# Quantum-classical correspondence in entanglement production: Entropy and classical tori

Shi-hui Zhang and Quan-lin Jie

*Department of Physics, Wuhan University, Wuhan 430072, P. R. China*

(Dated: January 18, 2008)

We analyze the connections between entanglement dynamics and classical trajectories in a semiclassical regime for two systems: A pair of coupled oscillators and the Jaynes-Cummings model. We find that entanglement production depends on classical invariant tori and such phenomenon is closely related to the power spectra of classical trajectories. Classical power spectrum with a larger number of frequency components corresponds to larger entanglement. We introduce a frequency entropy to describe the classical frequency distribution, and find that there is good correspondence between the classical frequency entropies and the maximum von Neumann entropies.

## I. INTRODUCTION

Problems of quantum-classical correspondence have been extensively investigated [1]. Recent studies have further shown the relations between entanglement production and the underlying classical dynamics for chaotic systems. In the early works on spin-boson system [2] and coupled kicked tops [3], it was suggested that entanglement production can be a good indicator of the regular to chaotic transition for classical chaotic systems. This claim has been affirmed in many subsequent studies [4,5,6]. Also it has been revised in some recent works [7-9]. An investigation on a pair of coupled kicked tops showed that increased chaos corresponds to a saturation of entanglement production rate in the weakly coupled regime [7]. Another study on Rydberg molecules [8] found that the quantum-classical correspondence in entanglement production is related to the inelastic scattering. Especially, a semiclassical approach shown by Jacquod [9] concluded that the entanglement production in bipartite system depends on both the degree of chaos and the coupling strength.

The aim of this paper is to further explore the connections between entanglement production and the underlying classical trajectories in a semiclassical regime for chaotic systems. In Section II, we investigate the dependence of entanglement production on the classical trajectories for the two coupled oscillators. We find that the production of entanglement is strongly related to the classical trajectories. The maximum values and rates of entanglement production correspond systematically to the classical invariant tori. When the initial state lies at the edge of regular islands for the regular case or in the chaotic sea for the mixed case, the production of entanglement is maximized.

Furthermore, the above correspondences between the production of entanglement and the classical trajectories are closely related to the classical power spectrum. The more complicated classical power spectra correspond to the larger maximum values and rates of entanglement production. We thus define a frequency entropy to measure the frequency distribution of classical power spectrum. It is shown that there is a good correspondence between the frequency entropies and the maximum von Neumann entropies for different classical trajectories (especially for regular ones). Such correspondence is also found in the investigations on the Jaynes-Cumming model presented in Section III. This is because a classical power spectrum reveals the structure of the energy levels involved in the quantum evolution [10,11,12]. The latter is directly connected to the entanglement production [13]. Thus the classical power spectrum and the classical frequency entropy can indicate the behavior of entanglement production.

## II. ENTANGLEMENT PRODUCTION IN A BIPARTITE MODEL

We now study the entanglement of a bipartite system with a model of two coupled oscillators, which is often used to analyze decoherence [14] or construct qubit models [15,16]. The Hamiltonian reads

$$H = \frac{p_1^2}{2m} + \frac{p_2^2}{2m} + \frac{1}{2}m\omega^2 q_1^2 + \frac{1}{2}m\omega^2 q_2^2 + \lambda q_1^2 q_2^2 , \quad (1)$$

where $m$ and $\omega$ are the mass and frequency of any of the two oscillators. $\lambda$ is the coupling parameter. This system is proposed by Pullen and Edmonds [17] and used in many works to study the classical-quantum correspondence for chaotic systems [18].

In this section, using the model (1), we investigate the entanglement dynamics in both the regular and mixed cases to achieve a detailed picture of the relations between the entanglement production and the classical trajectories. In the calculations, we adopt the natural units for the harmonic oscillator, i.e., $\hbar\omega$ for energy, $\sqrt{\hbar/m\omega}$ for length and $\sqrt{\hbar m\omega}$ for momentum. Accordingly, $\lambda$ in the model (1) is measured in units of $m^2\omega^3/\hbar$. The quantum dynamics is governed by the evolution operator $U(t) = \exp[-iHt/\hbar]$. The connection between quantum and classical domains is done by the preparation of the initial coherent state $|\psi(0)\rangle = |\alpha_1\rangle \otimes |\alpha_2\rangle$, whose center is precisely on a classical phase space point $(q_1, p_1; q_2, p_2)$ with $\alpha_l = (q_l + ip_l)/\sqrt{2}$ $(l = 1, 2)$ [19]. In the natural units, the state $|\alpha_l\rangle$ is a minimum uncertainty packet centered on a phase point with a width $O(1)$ in both the length and momentum directions. Besides, the entanglement is quantified by von Neumann entropy $S_V$ [20].

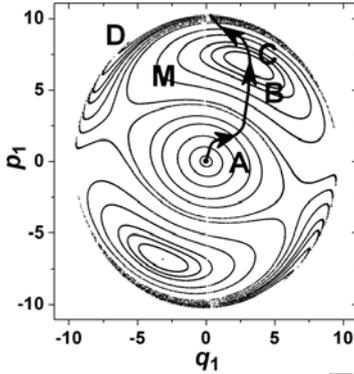

FIG. 1. Poincaré section ($q_2 = \sqrt{10}$ and $p_2 > 0$) for the regular case. The three regular islands are marked along the arrow line with **A**, **M** and **D**. **B** (**C**) denotes the upper (lower) half part of **M**. Unit of $q_1$ ($p_1$) is $\sqrt{\hbar/m\omega}$ ($\sqrt{\hbar m\omega}$).

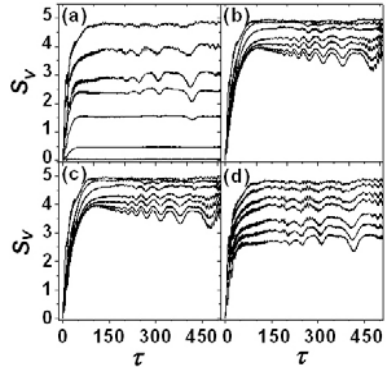

FIG. 2. von Neumann entropy $S_V$ vs time $\tau = \omega t$ for different initial states (see Table I). The CICS for the entropy curves in (a), (b), (c) and (d) are localized on the tori in **A**, **B**, **C** and **D** (see Fig. 1), respectively. Unit of $\tau$ is $\omega^{-1}$.

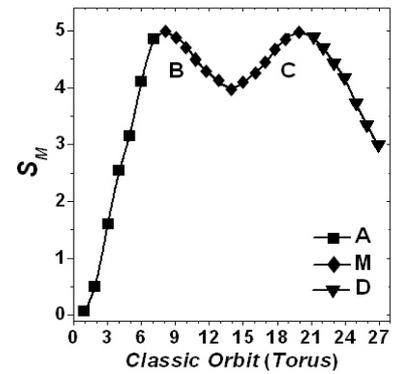

FIG. 3. Maximum von Neumann entropy $S_M$ vs corresponding classical trajectory. The squares, diamonds and triangles denotes the $S_M$'s corresponding to the tori in **A**, **M** and **D** (see Fig. 1), respectively.

### A. Entanglement production in the regular case

The classical dynamics of model (1) exhibits a regular phase space with $\lambda = 0.0075$ and total energy $E = 58$. In the classical phase space, we choose twenty one classical trajectories to explore the connections between entanglement production and classical dynamics. The Poincaré section of these trajectories is shown in Fig. 1. The three regular islands are marked with **A**, **M** and **D**. **B** (**C**) denotes the upper (lower) half part of the island **M**. In each regular island, there are seven tori.

From the tori in the island **A**, **M** and **D**, we pick up the centers of the initial coherent states (CICS) along an arrow line. As shown in Fig. 1, the arrow line starts from the origin in **A**, passes through the innermost torus in **M** and then reaches the minimum torus in **D**. Between the arrow line and the tori in **A**, **B**, **C** and **D**, there are twenty seven cross points. We pick up these cross points along the arrow line and display their coordinates in Table I in turn ($p_2$ are determined by energy conservation and thus are not shown hereafter). The rows A, B, C and D display the coordinates of the cross points sampled from the tori in **A**, **B**, **C** and **D**, respectively. For comparison purpose, the nodal coordinates taken from the innermost torus in **M** are shown in both the rows B and C with ($q_1/\sqrt{10}$, $p_1/\sqrt{10}$) = (1, 2.173).

The cross points chosen above are used as the CICS in the quantum evolution. With $E = 58$, the distance between any two adjacent cross points is approximately equal to 1 in the units used in this section. As mentioned above, the scales of the initial states in the same units are $O(1)$ in both the length and momentum directions. Thus, the distance between any two adjacent CICS is approximately equal to the scales of the initial states. This makes sure that we are treating a semiclassical regime.

With the CICS chosen above, we calculate the von Neumann entropies and show the results in Fig. 2. The entropy curves in Fig. 2(a) [Fig. 2(c)] from bottom to top correspond to the CICS in the row A (C) of Table I from first to last, while those in Fig. 2(b) [Fig. 2(d)] from top to bottom correspondent to the CICS in the row B (D) from first to last. Accordingly, the entropy curves in Figs. 2(a), 2(b), 2(c) and 2(d) correspond to the tori in **A**, **B**, **C** and **D** in Fig. 1, respectively. Especially, as the tori in **A**, **B**, **C** and **D** enlarge from the innermost to the outermost, the entropy curves in the corresponding subfigure of Fig. 2 increase from bottom to top.

TABLE I. Coordinates of the centers of the initial coherent states (CICS) sampled from the tori in Fig. 1

| Regions | ($q_1/\sqrt{10}$, $p_1/\sqrt{10}$) with fixed $q_2/\sqrt{10} = 1$ and $p_2 > 0$ |
|---|---|
| **A** | (0, 0) , (0, 0.121), (2/9, 0.242), (4/9, 0.36), (1, 0), (1, 0.458), (1, 0.801); |
| **B** | (0.74,1.08), (0.9,1.31), (0.92,1.53), (0.96,1.74), (0.98, 1.84), (1, 1.95), (1, 2.173); |
| **C** | (1,2.173), (0.9, 2.4072), (0.8,2.5234), (0.6266,2.631), (0.33,2.8016), (0.245,2.933), (0.126,3.0748); |
| **D** | (0.1,3.11), (0.07,3.14), (0.06,3.17), (0.04,3.187), (0.02,3.21), (0.01,3.2,32), (0,3.25576) |

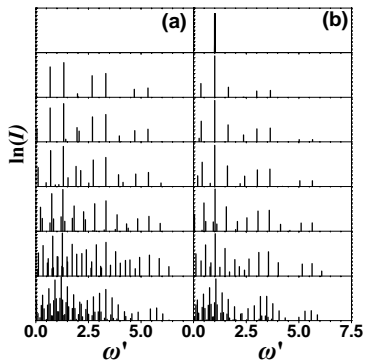

FIG. 4. Power spectra of (a) $q_1$ and (b) $q_2$ for the trajectories whose tori lie in the island **A** (see Fig.1). The seven pairs of spectra from top to bottom correspond to the seven tori in **A** from the innermost to the outermost. Unit of $\omega'$ is $\omega$.

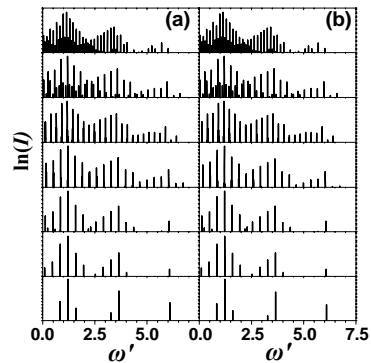

FIG. 5. As in Fig. 4 but for the trajectories whose tori lie in the island **M** (see Fig.1). The seven pairs of spectra from top to bottom correspond to the seven tori in **B** or **C** from the innermost to the outermost. Unit of $\omega'$ is $\omega$.

To clarify the dependence of entanglement production on classical trajectories, we further calculate the maximum values of the entropy curves in Fig. 2 and show the results (i.e., the maximum von Neumann entropies) in Fig. 3. The maximum entropies $S_M$ in Fig. 3 are arranged in the same order as their corresponding CICS are picked up from the tori along the arrow line in Fig. 1 (i.e., the CICS of the $k$ th $S_M$ in Fig. 3 correspond to the $k$ th cross point picked up from the tori along the arrow line in Fig. 1).

In Fig. 3, the maximum von Neumann entropies $S_M$ vary significantly with the corresponding classical invariant tori. For instance, as the tori in the island **A** enlarge from the innermost to the outermost, the corresponding $S_M$'s in Fig. 3 increase from the first square to the seventh square. Then, as the tori in **B** shrink from the outermost to the innermost, the values of $S_M$ in Fig. 3 decrease from the first diamond to the valley between 'B' and 'C'. For the CICS localized on the tori in **C** and **D**, the curve of $S_M$ also varies with the classical tori in a similar way. Roughly speaking, the value of $S_M$ increases (decreases) with the enlargement (shrink) of the classical tori and is maximized when the corresponding torus lies at the edge of the regular islands. Furthermore, since different entropy curves in Fig. 2 reach the saturation plateaus almost at the same time, a larger $S_M$ indicates a faster generation of entanglement. Thus, the rate of entanglement production also roughly increases (decreases) with the enlargement (shrink) of the classical tori.

Besides, in the investigations, we also note that there is good correspondence between the oscillations of the entropy $S_V(t)$ and those of the underlying classical coupling potential $\lambda q_1^2(t) q_2^2(t)$ for some classical trajectories. The oscillations of classical coupling potential reveal the frequency distribution of classical trajectory. We thus perform frequency analyses of the classical trajectories and further find interesting connections between the en-tanglement production and the classical power spectrum.

Classical power spectrum is used to analyze the frequencies of classical motion [21]. For a classical trajectory, the classical motion is generally multiply periodic, admitting the Fourier expansion [22]

$$\mathbf{x}(t) = \sum_{\mathbf{m}} \mathbf{X}_{\mathbf{m}}(\mathbf{I}) \exp\left[i\mathbf{m}\cdot(\boldsymbol{\omega} t + \boldsymbol{\theta}_0)\right], \quad (2)$$

where $\mathbf{x}(t)$ stands for some dynamical variable such as position coordinate $\mathbf{q}(t)$, and $\boldsymbol{\omega}$ are the frequencies related to the actions $\mathbf{I}$. The integer set $\mathbf{m}$ $(m_1, m_2)$ goes from $-\infty$ to $+\infty$ and the coefficients $\mathbf{X}_{\mathbf{m}}(\mathbf{I})$ are dependent on $\mathbf{I}$. With the Fourier expansion (2), the frequencies of a classical trajectory can be obtained using a numerical integration [22]

$$I_\alpha(\omega') = \frac{1}{2\pi} \lim_{T \to \infty} \frac{1}{T} \left| \int_0^T dt\, x_\alpha(t) \exp(-i\omega' t) \right|^2 \quad (3)$$

where $x_\alpha(t)$ denotes the $\alpha$-th component of $\mathbf{x}(t)$. Specifically, for the $\alpha$th component of $\mathbf{q}(t)$ $(\alpha = 1, 2)$, $q_\alpha(t) = \sum_{\mathbf{m}} Q_{\mathbf{m}}^\alpha(\mathbf{I}) \exp[i\,\mathbf{m}\cdot(\boldsymbol{\omega} t + \boldsymbol{\theta}_0)]$ and the power spectrum becomes $I_\alpha(\omega') = \sum_{\mathbf{m}} |Q_{\mathbf{m}}^\alpha|^2 \delta(\omega' - \mathbf{m}\cdot\boldsymbol{\omega})$.

With the EBK method shown in Ref. [21], we calculate the classical power spectra of $q_\alpha(t)$ $(\alpha = 1, 2)$ for the twenty one classical trajectories chosen above (the evolution time $\tau = \omega t = 2^{17}$). Figures 4 and 5 display the results for the trajectories whose tori lie in **A** and **B** (see Fig. 1), respectively. The results for the tori in **D** (**C**) are similar to those for the tori in **A** (**B**) and thus are not shown here. For comparison purpose, the strength axis in the power spectrum is plotted in logarithmic scale (i.e., $\ln[I(\omega')]$ hereafter).

The seven pairs of power spectra in Fig. 4 from top to bottom correspond to the seven entropy curves in Fig. 2(a) from bottom to top, while those in Fig. 5 from top to bottom correspond to the seven $S_V(t)$ curves in Fig. 2(b) from top to bottom. It is clear that more complicated power spectra correspond to higher maximum values and faster rates of entanglement production. Such correspondence is also found in the investigations related to the tori in **C** and **D**.

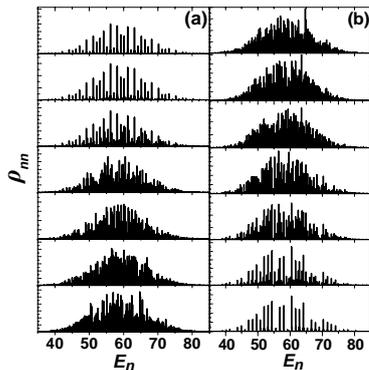

FIG. 6. Density spectrum $\{\rho_{nn}\}$ vs corresponding eigenenergy $E_n$. The spectra in the column (a) [(b)] from top to bottom correspond to the CICS in the row A (B) of Table I from first to last. Unit of $E_n$ is $\hbar\omega$.

In the spectra in Figs. 4 and 5, there are sharp lines at the frequencies $\omega'$. Each line corresponds to a transition from one energy level to another. Its strength is related to a physical process, which involves the motion of the two particles [22]. Thus, classical power spectrum indicates the energy levels involved in the corresponding quantum evolution. The latter can be revealed by quantum density spectrum.

Quantum density spectrum is expressed as $\rho_{nn} = \langle n | \rho(t) | n \rangle$ [13], where $\rho(t)$ is the quantum density at time $t$ and $|n\rangle$ are the eigenvectors of the Hamiltonian with $E_n$ as the corresponding eigenvalues. According to Ghose *et al.* [13], a larger set of $\{\rho_{nn}\}$ correspond to a faster generation of entanglement in the regular to chaotic transition. Figures 6(a) and 6(b) display the quantum density spectra for the CICS localized on the tori in **A** and **B**, respectively. The seven spectra in Fig. 6(a) from top to bottom correspond to the seven pairs of classical power spectra in Fig. 4 from top to bottom. The same correspondence also holds between the quantum density spectra in Fig. 6(b) and the classical power spectra in Fig. 5. From the above comparisons, we can see that more complicated classical power spectrum corresponds to larger set of $\{\rho_{nn}\}$. Since $\{\rho_{nn}\}$ is closely related to quantum entropy [13], the correspondence between the classical power spectrum and the quantum density spectrum indicates the connections between the entanglement production and the classical trajectories.

## B. Entanglement production in the mixed case

With $\lambda = 0.0075$ and $E = 150.75$, the classical dynamics of the Hamiltonian (1) exhibits a mixed phase space. In the classical phase space, we choose seventeen regular classical trajectories and four chaotic ones, and show their Poincaré section in Fig. 7. The three islands formed by the regular trajectories are marked along the arrow line with **A**, **M** and **D**. **B** (**C**) denotes the lower (half) part of the island **M**. In each regular island, there are six tori.

Between the arrow line and the tori in **A**, **M** and **D**, there are twenty three cross points. These cross points are used as the CICS in the quantum evolution. Along the direction of the arrow line in Fig. 7, we pick up these cross points and present their coordinates in turn in the lower four rows of Table II. Specifically, the rows A, B, C and D show the CICS sampled from the tori in **A**, **B**, **C** and **D** (see Fig. 7), respectively. For comparison purpose, the coordinates of the CICS sampled from the innermost torus in **M** are shown in both the rows B and C with $(q_1/\sqrt{10}, p_1/\sqrt{10}) = (1/4, 3.8737)$. Besides, we sample four CICS from the chaotic sea and display their coordinates in the first row of Table II. Similar to the regular case, with $E = 150.75$, the distance between any two CICS in Table II is also approximately equal to or larger than the length and momentum scales of the initial states. In addition, the initial states corresponding to the chaotic CICS are also fully contained in the chaotic sea.

We calculate the von Neumann entropies for the CICS in Table II, and present the results in Fig. 8. The topmost entropy curves in Figs. 8(a) to 8(d) correspond to the four chaotic CICS in Table II from first to last. The lower six entropy curves in Fig. 8(a) [Fig. 8(c)] from bottom to top correspond to the six CICS in the row A (B) of Table II from first to last, while the lower six entropy curves in Fig. 8(b) [Fig. 8(d)] from top to bottom correspond to the six CICS in the row B (D) of Table II from first to last.

We further calculate the maximum values $S_M$ of the entropy curves in Fig. 8 and display the results in Fig. 9. The four $S_M$'s marked with circles from first to last correspond to the four chaotic CICS in Table II from first to last. The $S_M$'s marked with squares, diamonds and trian-gles correspond to the regular CICS localized on the tori in the island **A**, **M** and **D**, respectively. Similar to the regular case, the $S_M$'s obtained with the regular CICS are arranged in Fig. 9 in the same order as their corresponding CICS are sampled from the tori along the arrow line in Fig. 7. For instance, the CICS of the $S_M$'s numbered from one to six in Fig. 9 are sampled from the six tori in **A** from the innermost to the outermost. In Fig. 9, the curve of $S_M$ varies with the classical tori and is maximized when the CICS lies in the chaotic sea. So do the growth rates of the entropy curves as shown in Fig. 8.

TABLE II. Coordinates of the chaotic CICS and those of the regular CICS sampled from the tori in Fig. 7

| Regions | $(q_1/\sqrt{10}, p_1/\sqrt{10})$ with fixed $q_2/\sqrt{10} = 1/4$ and $p_2 > 0$ |
|---|---|
| Chaotic sea | (1/4, 2.2427), (-2.3963, 1.3352), (-2.0246, -1.1064), (1/4, 4.4846); |
| **A** | (0, 0), (1/36, 0.2041), (1/4,0), (1/4, 0.7155), (1/4, 1.5050), (1/4, 1.9102); |
| **B** | (1/4, 2.9143), (1/4, 3.0582), (1/4, 3.2621), (1/4, 3.4660), (1/4, 3.6699), (1/4, 3.8737); |
| **C** | (1/4, 3.8737), (1/4, 4.0690), (1/4, 4.2440), (1/4, 4.4026), (1/4, 4.5466), (1/4, 4.6402); |
| **D** | (1/4, 5.1357), (1/4, 5.2687), (1/4, 5.3412), (1/4, 5.4184), (1/4, 5.4643), (1/4, 5.4795). |

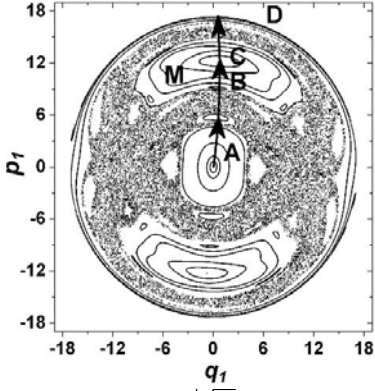

FIG. 7. Poincaré section ($q_2/\sqrt{10}=1/4$ and $p_2>0$) for the classical trajectories chosen in the mixed case. The three regular islands are marked along the arrow line with **A**, **M** and **D**. **B** (**C**) denotes the upper (lower) half part of **M**. Unit of $q_1$ is $\sqrt{\hbar/m\omega}$, and that of $p_1$ is $\sqrt{\hbar m\omega}$.

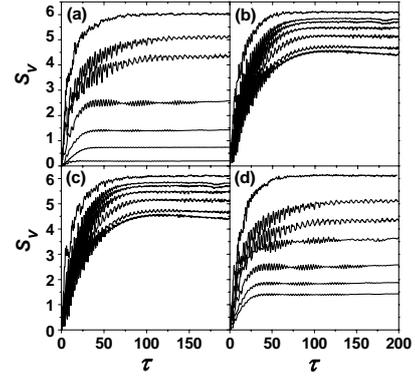

FIG. 8. von Neumann entropy $S_V$ vs time $\tau=\omega t$ for the CICS on different classical trajectories (see Table II). The top-most curves in (a) to (d) correspond to the four chaotic CICS. The others in (a), (b), (c) and (d) correspond to the CICS on the tori in **A**, **B**, **C** and **D**, respectively. Unit of $\tau$ is $\omega^{-1}$.

We investigate the power spectra of all the above classical trajectories with the EBK methods (the evolution time $\tau=2^{17}$) [21]. In Fig. 10, we exhibits the results for the six tori in the island **A** and those for a chaotic trajectory with ($q_1/\sqrt{10}, p_1/\sqrt{10}$)=(1/4, 2.2427) (see Table II) as initial conditions. The upper six pairs of spectra in Fig. 10 from top to bottom correspond to the six tori in the island **A** from the innermost to the outermost. Also they correspond, from top to bottom, to the lower six entropy curves in Fig. 8(a) from bottom to top. The bottom pair is the power spectra of the chaotic trajectory and corresponds to the top entropy curve in Fig. 8(a). A comparison between Figs. 10 and 8(a) shows that more complicated power spectra correspond to larger and faster entanglement production. Such correspondence is also found in the investigations on the relation between the entanglement production and the classical power spectra for the other classical trajectories. Especially, for the chaotic trajectory, the power spectrum is continuous and the corresponding entanglement production is maximized.

In Fig. 8, there are relatively small differences in both the maximum values and the growth rates between the four chaotic entropy curves. This is because the instabilities of the chaotic trajectories are similar to each other, for the chaotic power spectra are all continuous and the Lyapunov exponents of the four chaotic trajectories are around $0.12\pm0.02$. This is in accord with the previous suggestion [3,9,23] that the enhancement of entanglement production induced by chaos depends on the Lyapunov exponent.

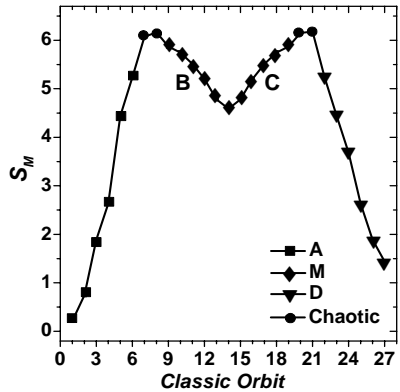

FIG. 9. Maximum von Neumann entropy $S_M$ vs corresponding classical trajectory. The circles correspond to the chaotic trajectories. The squares, diamonds and triangles correspond to the tori in **A**, **M** and **D**, respectively.

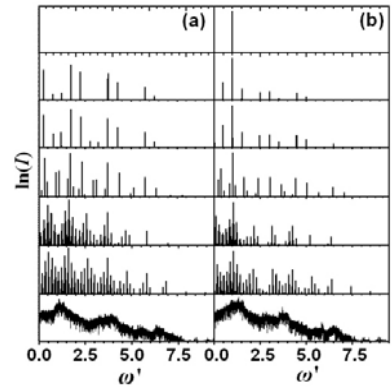

FIG. 10. Power spectra of (a) $q_1$ and (b) $q_2$ for different classical trajectories. The upper six pairs from top to bottom correspond to the six tori in **A** (Fig. 7) from the innermost to the outermost. The bottom pair corresponds to a chaotic trajectory. Unit of $\omega'$ is $\omega$.

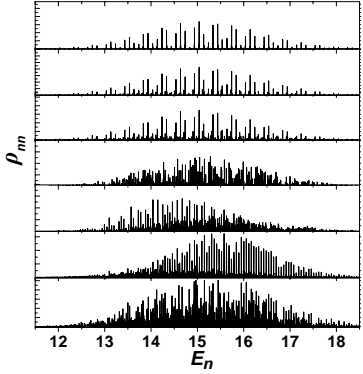

FIG. 11. Density spectra $\{\rho_{nn}\}$ vs corresponding eigenenergy $E_n$. The upper six spectra from top to bottom correspond to the six CICS in the row A of Table II from fist to last. The bottom one correspond to the CICS localized in the chaotic sea with ( $q_1/\sqrt{10}$, $p_1/\sqrt{10}$ )=(1/4, 2.2427 ). Unit of $E_n$ is $\hbar\omega$ .

A comparison of Figs. 3 and 9 shows that the curve of $S_M$ in Fig. 3 is similar to that in Fig. 9, though the classical dynamics transits from the regular case to the mixed case. Especially, the variation of $S_M$ in Fig. 9 through the transition to the chaotic trajectories is similar to that in Fig. 3 through the transition to the edges of the regular islands. This suggests that entanglement production is an indicator not only of the regular to chaotic transition but also of the transitions in different regular classical trajectories.

To further investigate the quantum-classical correspondence in entanglement production, we calculate the quantum density spectra corresponding to the CICS shown in Table II. In Fig. 11, we exhibit the results for the CICS shown in the row A of Table II and those for a chaotic CICS with ( $q_1/\sqrt{10}$, $p_1/\sqrt{10}$ )=(1/4, 2.2427). The seven density spectra in Fig. 11 from top to bottom correspond to the seven entropy curves in Fig. 8(a) from bottom to top. Also they correspond, from top to bottom, to the classical power spectra in Fig. 10 from top to bottom. Similar to the regular case, a comparison of Figs. 10 and 11 reveals that a larger set of frequency components involved in the classical dynamics corresponds to a larger set of eigenstates involved in the quantum evolution.

## C. Frequency entropy and quantum-classical correspondence

In the above investigations on the coupled oscillators, we show that entanglement production depends systematically on classical invariant tori. When the initial state lies at the edge of regular islands for the regular case or in the chaotic sea for the mixed case, the entanglement production is maximized. As shown in the previous studies [3, 9, 23], the enhancement of entanglement production induced by chaos has been investigated and explained by the methods related to the Lyapunov exponents.

Here we further develop a practical method to describe the dependence of entanglement production on classical tori from the perspective of classical power spectra. We use $I_k^\alpha(\omega')$ to stand for the amplitude of the $k$ th component in the discrete power spectrum of $q_\alpha$ ( $\alpha=1,2$ ). To describe the discrete classical spectral distribution, we consider a new quantity as

$$J_k^\alpha(\omega') = I_k^\alpha(\omega')\Big/\sum\nolimits_{\alpha,k} I_k^\alpha(\omega') .\qquad(4)$$

We then define a frequency entropy $S_{Fr}$ as a measure of the classical frequencies, i.e.,

$$S_{Fr} = -\sum\nolimits_{\alpha,k} J_k^\alpha(\omega')\ln J_k^\alpha(\omega') ,\qquad(5)$$

which is used to distinguish different regular trajectories and measure the stability of classical dynamics.

With the above method, we investigate the classical frequency entropies of all the regular trajectories chosen in the regular and mixed cases. The results are presented in Fig. 12. The classical frequency entropies $S_{Fr}$ and the corresponding maximum von Neumann entropies $S_M$ are represented by squares and circles, respectively. The $S_M$'s are shown in the same way in Fig. 12(a) [Fig. 12(b)] as they are shown in Fig. 3 (Fig. 9). In Figs. 12(a) and 12(b), there is good correspondence between the variations of $S_M$ and $S_{Fr}$ with the classical trajectories. This clearly indicates the quantum-classical correspondence in the production of entanglement.

For the continuous power spectrum of chaotic trajectories, the summation over $k$ in Eqs. (4) and (5) is re-placed by the integral over the frequency $\omega'$. Using this method, we can obtain the frequency entropies of classical chaotic trajectories. However, due to the continuity of chaotic power spectrum, the classical frequency entropies of chaotic trajectories are not consistent with those of regular trajectories. Further research will be needed to study the classical frequency entropies of chaotic trajectories. It is conceivable, however, that in Fig. 9 the $S_M$'s corresponding to the chaotic trajectories are larger than those corresponding to the regular trajectories, for the chaotic power spectra are continuous and have more frequency components.

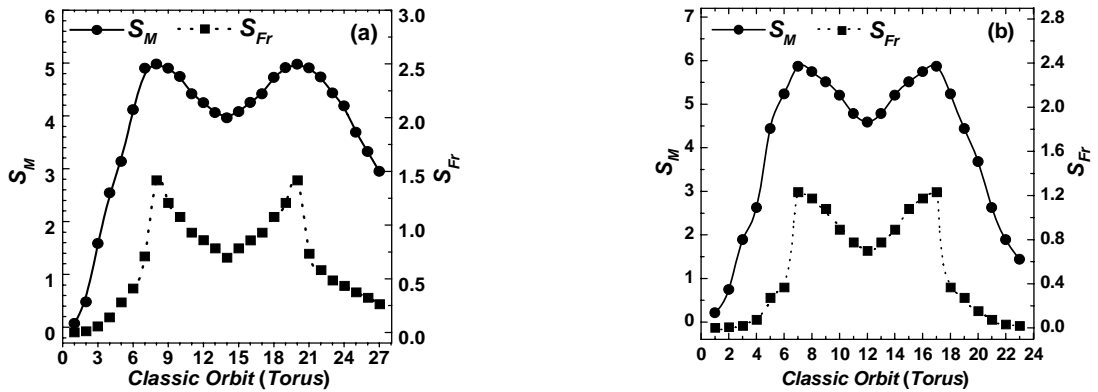

FIG. 12. Maximum von Neumann entropy $S_M$ (solid) and classical frequency entropy $S_{Fr}$ (dotted) vs corresponding regular trajectory in (a) the regular case ( $E=58$ ) and (b) the mixed case ( $E=150.75$ ).

As mentioned above, the enhancement of entanglement induced by chaos has been discussed and explained by the methods related to the Lyapunov exponent. Here, to further understand the connections between the production of entanglement and the power spectrum of regular classical trajectories, we can get some clews from a general method of quantum-classical correspondence presented by P. Brumer *et al.* based on the Liouville mechanics [10,11,12]. Following this method, the time evolution of a classical property $\mathbf{x}(t)$ can be expressed as [10,12]

$$\mathbf{x}(t) = tr[\rho(t)\mathbf{x}] = \sum_{\lambda,\alpha} \mathbf{X}_{\lambda,\alpha} \exp[-i\lambda t], \quad (6)$$

where $\mathbf{x}(t)$ denotes some dynamical variable [e.g., position coordinate $\mathbf{q}(t)$] and $\rho(t)$ is the classical density function. The coefficient $\mathbf{X}_{\lambda,\alpha}$ depends on the initial conditions. The parameter $\lambda$ is the classical correspondence of $\lambda_{\mathbf{nm}} = (E_{\mathbf{m}} - E_{\mathbf{n}})/\hbar$ [12], where $E_{\mathbf{n}}(E_{\mathbf{m}})$ is the eigenvalue of the Hamiltonian with $|\mathbf{n}\rangle(|\mathbf{m}\rangle)$ as the corresponding eigenstate. The label $\alpha$ is the degeneracy related to $\lambda$ [12].

For regular motion, $\lambda = \mathbf{k}\cdot\boldsymbol{\omega}(\mathbf{I}')$[12], where each member of the integer set $\mathbf{k}$ goes from $-\infty$ to $+\infty$ and $\boldsymbol{\omega}(\mathbf{I}')$ is the frequency dependent on the action $\mathbf{I}'$. By comparing Eq. (6) with Eq. (2), one can see that the set of frequencies $\{\omega' = \mathbf{m}\cdot\boldsymbol{\omega}\}$ in Eq. (2) are equivalent to $\{\lambda = \mathbf{k}\cdot\boldsymbol{\omega}(\mathbf{I}')\}$ in Eq. (6) (each member of the set $\mathbf{m}$ also goes from $-\infty$ to $+\infty$). Since $\lambda = \mathbf{k}\cdot\boldsymbol{\omega}(\mathbf{I}')$ corresponds to $\lambda_{\mathbf{nm}} = (E_{\mathbf{m}} - E_{\mathbf{n}})/\hbar$ in the quantum counterpart, the frequency lines at $\omega' = \mathbf{m}\cdot\boldsymbol{\omega}$ in the classical power spectrum reveal the structure of the energy levels involved in the quantum evolution. More complicated classical power spectrum indicates larger set of eigenstates extending over broader and denser quantum density spectrum (the latter can enhance the entanglement production [13]). Such connection is also revealed in the comparisons between the classical power spectra and the quantum density spectra shown above. Similar to the Lyapunov exponent, the classical power spectrum and the classical frequency entropy suggest a simple method to indicate the dependence of entanglement on classical dynamics.

## III. ENTANLEMENT PRODUCTION OF THE JAYNES-CUMMINGS MODEL

To confirm the connection between the entanglement and the classical frequency entropy, we perform similar investigations on the Jaynes-Cummings model. The Hamiltonian reads [2]

$$H = \hbar\omega a^\dagger a + \varepsilon J_z + \frac{G}{\sqrt{2J}}\left(aJ_+ + a^\dagger J_-\right) + \frac{G'}{\sqrt{2J}}\left(a^\dagger J_+ + aJ_-\right), (7)$$

where the first term corresponds to the free single-mode field with frequency $\omega$, the second term corresponds to the $N = 2J$ two-level atoms with energy separation $\hbar\varepsilon$ and the other terms are the interaction between the field and the atoms. Here, we set $\omega = \varepsilon = 1$ and use the natural units, i.e., $\hbar = c = 1$.

The connection between the classical and quantum domains is done by the choice of the initial states. Given a point $(q_1, p_1; q_2, p_2)$ in the classical phase space, the corresponding quantum initial state is $|\psi(0)\rangle = |w\rangle \otimes |v\rangle$. $|w\rangle(|v\rangle)$ are the atomic (field)-coherent states given by $|w\rangle = (1 + w\bar{w})^{-J} e^{wJ_+} |J, -J\rangle$ and $|v\rangle = e^{-w\bar{w}/2} e^{va^\dagger} |0\rangle$ with $w = (p_1 + iq_1)/\sqrt{4J - (p_1^2 + q_1^2)}$ and $v = (p_2 + iq_2)/\sqrt{2}$ [24].

The classical version of the Hamiltonian (7) is obtained by a procedure as $H(v, v^*; w, w^*) \equiv \langle wv|H|wv\rangle$. It can be rewritten in terms of the phase-space variables as [25]

$$H(q_1, p_1; q_2, p_2) = \frac{\omega}{2}(q_2^2 + p_2^2) + \frac{\varepsilon}{2}(q_1^2 + p_1^2 - 2J) + \sqrt{1 - (q_1^2 + p_1^2)/4J}(G_+ p_1 p_2 + G_- q_1 q_2)$$

$$, (8)$$

where $G_\pm = G \pm G'$.

Using the model (7), we investigate the connections between the von Neumann entropies and the frequency entropies for different classical trajectories in both regular and mixed cases. In the calculations, the evolution time for calculating $S_M$ is approximately equal to 350; and that for calculating $S_{Fr}$ is approximately equal to $5\times10^5$ (in the natural units).

The Poincaré section in Fig. 13 exhibits a regular phase space of the model (8) with $E = 40$. As shown in Fig. 13, we pick up fifteen CICS along the arrow line from different tori and present their coordinates $(q_1, p_1)$ in turn as follows: (0.1,-5.005), (0.1,-4.6046), (0.1,-4.2042), (0.1, -3.8038), (0.1, -3.4034), (0.1, -2.6026), (0.1, -1.8018) and (0.1, -0.6006), from tori in the lower island; (0.1,1.8018), (0.1, 2.6026), (0.1, 3.1534), (0.1,3.5538), (0.1,4.9542), (0.1, 4.3546) and (0.1, 5.005), from the tori in the upper island. For all the CICS in this section, $q_2$=0, $p_2 > 0$ and the values of $p_2$ are determined by energy conservation.

Figure 14 shows the maximum von Neumann entropies $S_M$ (solid line) for the above fifteen CICS and their corresponding classical frequency entropies $S_{Fr}$ (dotted line). In Fig. 14, the $S_M$'s numbered from one to fifteen correspond to the above fifteen CICS from first to last. As shown in Fig. 14, the variations of $S_M$ and $S_{Fr}$ with the classical tori are synchronous.

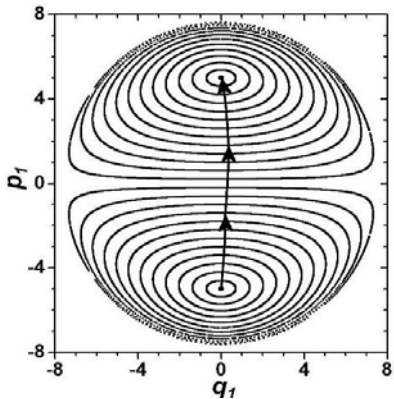

FIG. 13. Poincaré section ($q_2 = 0$, $p_2 > 0$) for the Jaynes-Cummings model with $E = 40$, $J/2 = 14.5$, $G = 0.25$ and $G' = 0$. $q_1$ and $p_1$ are given in the natural units.

Figure 15 shows the Poincaré section of the classical Hamiltonian (8) in a mixed case. We sample the CICS from different classical trajectories along the arrow line and show their coordinates ($q_1, p_1$) in turn as follows: (0.1, -5.86), (0.1,-5.0792), (0.1,-4.27) and (0.1,-3.4521), from the tori in the island **A**; (2.21,-2.2924) and (2.424, -3.078), from the chaotic sea; (4.21, -2.19), (4.9,-1.46), (4.9, -0.8), (4.7471, -0.4992) and (4.7295, 0.13558), from the tori in the island **B**; (2.2302, -0.66822), from the chaotic sea; (1.71259, 2.6883), (1.0029, 3.2417), (0.1,3.5096) and (0.1, 3.9), from the tori in the island **C**.

We calculate the maximum von Neumann entropies $S_M$ for the above sixteen CICS and show the results in Fig. 16. Here, the $S_M$'s from first to last correspond to the above sixteen CICS from first to last. The curve of $S_M$ in Fig. 16 exhibits the variation of the maximum entanglement with the underlying classical trajectories. Among these classical trajectories, there are thirteen regular ones, whose tori lie in the island **A**, **B** and **C** (Fig. 15). Figure 17 shows the classical frequency entropies for these regular classical trajectories. In Fig. 17, the squares and circles stand for the classical frequency entropies and their corresponding maximum von Neumann entropies, respectively. From Fig. 17, one also can see that there is good correspondence between the maximum von Neumann entropies and the classical frequency entropies.

## IV. CONCLUSION

In this paper, using the model of coupled oscillators and the Jaynes-Cummings system, we show that the entanglement production of bipartite system changes systematically with classical invariant tori in both the regular and mixed cases. Furthermore, the dependence of entanglement on the underlying classical trajectories is closely related to the classical power spectrum. The larger the number of classical frequency components is, the larger the corresponding entanglement is. Such phenomena reveal the quantum-classical correspondence in the entanglement production from the perspective of classical trajectories. This sheds lights on the further understanding of the quantum-classical correspondence in the entanglement production of chaotic systems.

## ACKNOWLEDGEMENTS

This work is supported by the National Science Foundation (Grant No. 1037504).

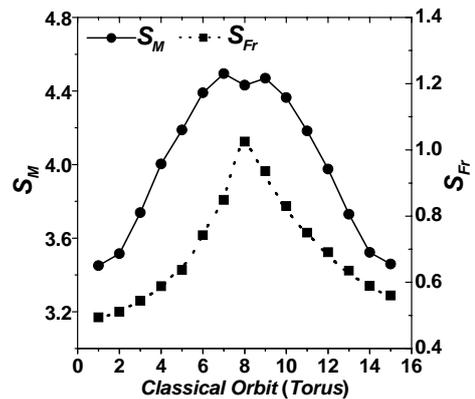

FIG. 14. Maximum von Neumann entropy $S_M$ (solid) and classical frequency entropy $S_{Fr}$(dotted) vs corresponding torus (see Fig. 13).

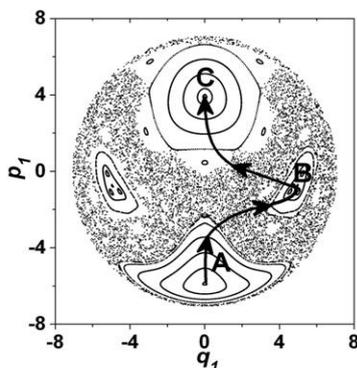

FIG. 15. Poincaré section ($q_2 = 0$ and $p_2 > 0$) for the Jaynes-Cummings model with $E = 35$, $J/2 = 12.5$, $G = 0.4$ and $G' = 0.25$. The three islands are marked along the arrow line with **A**, **B** and **C**. Units are as in Fig. 13.

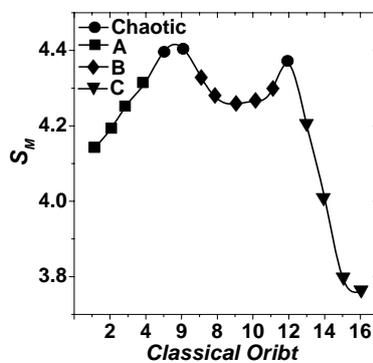

FIG. 16. Maximum von Neumann entropy $S_M$ vs corresponding classical trajectory. The circles correspond to the chaotic trajectories. The squares, diamonds and triangles correspond to the tori in **A**, **B** and **C**, respectively.

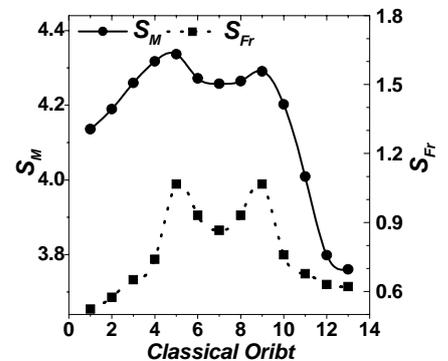

FIG. 17. Maximum von Neumann entropy $S_M$ (solid) and classical frequency entropy $S_{Fr}$ (dotted) for the regular trajectories vs corresponding tori shown in Fig. 15.